\documentclass[12pt,a4paper]{article}
\pdfoutput=1
\usepackage{amsmath,amssymb,exscale}
\usepackage{amsfonts,amsthm}
\usepackage{graphicx, caption, float, epsfig}
\usepackage{mathrsfs}
\usepackage{slashed}
\usepackage{textcomp}
\usepackage[T1]{fontenc}
\usepackage[utf8]{inputenc}
\usepackage[table]{xcolor}
\usepackage[lowtilde]{url}
\usepackage{cancel}
\usepackage[font={small}]{caption}
\numberwithin{equation}{section}
\usepackage{verbatim} 
\usepackage{appendix}
\usepackage{hyperref}

\newcommand{\be}{\begin{equation}}
\newcommand{\ee}{\end{equation}}
\newcommand{\ba}{\begin{eqnarray}}
\newcommand{\ea}{\end{eqnarray}}
\newcommand{\bay}{\begin{array}{rcl}}
\newcommand{\eay}{\end{array}}
\newcommand{\ra}{\rightarrow}

\begin{document}

\title{The fate of supersymmetry in quantum field theories}

\author{Risto Raitio\footnote{E-mail: risto.raitio@helsinki.fi}\\Helsinki Institute of Physics, P.O. Box 64, \\00014 University of Helsinki, Finland}

\date{January 29, 2024}  \maketitle 

\abstract{\noindent
We analyze the significance of supersymmetry in two topological models and the standard model (SM). We conclude that the two topological field theory models favor hidden supersymmetry. The SM superpartners, instead, have not been found.}

\vskip 3cm


\newpage
\tableofcontents
\vskip 2.0cm

\section{Introduction}
\label{intro}

Matter in two topological field theory scenarios goes through one or two phase transitions between Planck time and the present time. We analyze these two models to determine what happens to supersymmetry (SUSY) at laboratory energies provided it is valid, say, at the grand unified (GUT) scale. The point of this note is to provide evidence that two attitudes, no supersymmetry and very heavy superpartners, are not justifiable in the light of present experimental measurements. For the standard model our argument is based on improved coupling constant behavior in grand unified theories.

The article is organized as follows. In section \ref{earlyuniv} we consider some general features, like the three different phases of the universe, the phase transitions and motivation for preons (called here chernons). To indicate the nature of problem of phase I matter, two models of topological gravity are briefly reviewed in section \ref{topolgrav}. Comparison of the present scenario and standard model inflation is made in section \ref{compar}. Conclusions and outlook are given in section \ref{conclusions}. An appendix with table \ref{tab:table2} of CS particle - SM particle correspondence is provided.

\section{The phases of the evolving universe}
\label{earlyuniv}

\begin{figure}[H]
	\centering
	\captionsetup{width=.8\linewidth}
	\includegraphics[width=10cm]{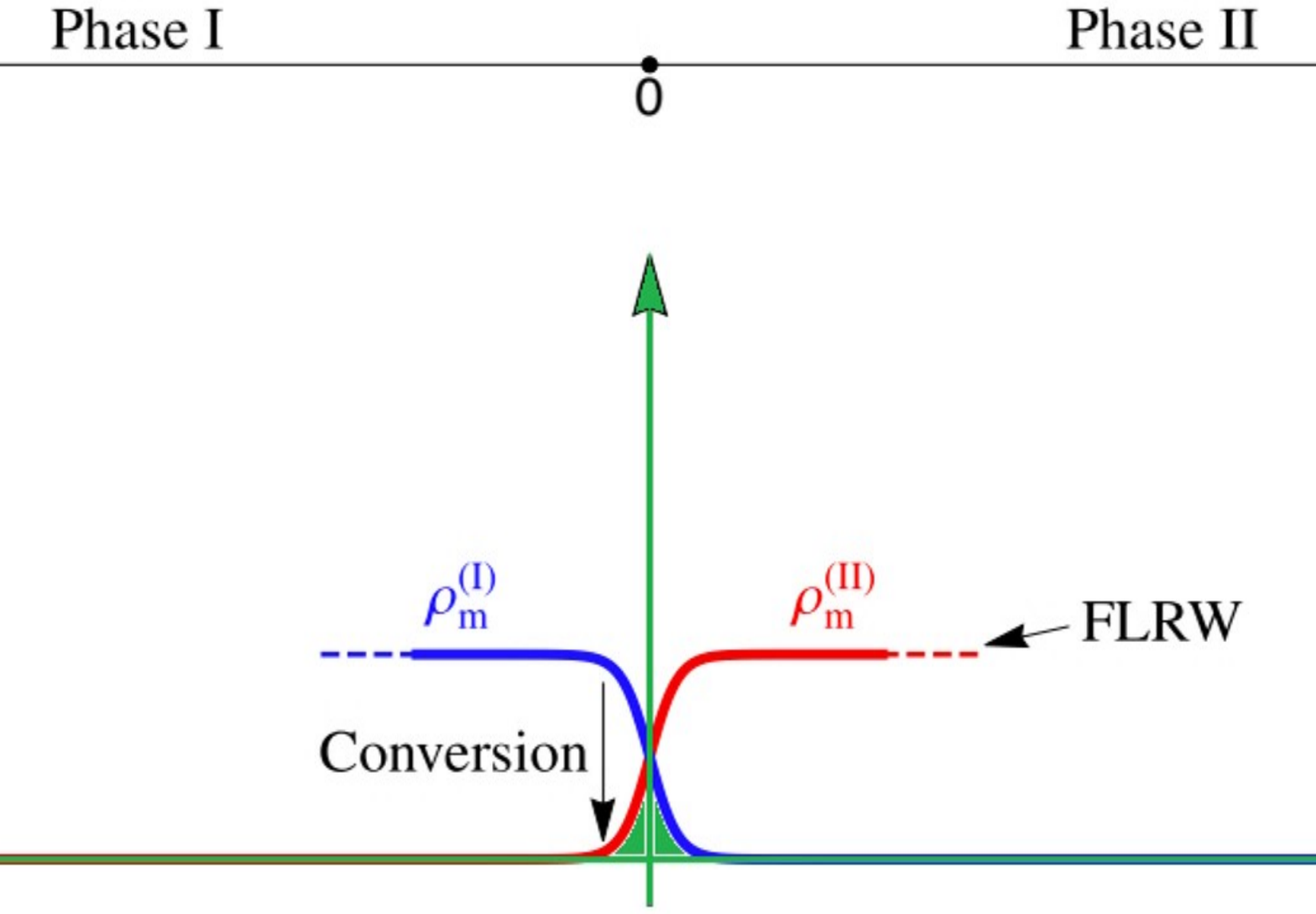}
	\caption{\small Transition from phase I to II in the universe proceeds by the conversion of matter made up from the degrees of freedom of frame (blue) to those of our T-dual frame (red). In our model the time t = 0 corresponds to energy scale $\Lambda_{cr}$. The small green area around $t \sim 0$ is the new second topological phase O, to be discussed in subsection \ref{chernon}. In the other topological model the point t = 0 is near Planck scale. In the SM t = 0 corresponds the GUT scale. (Figures 1, 2, and 4 are from \cite{Vafa&al} with permission.)}
	\label{fig:figure1}
\end{figure}

The common view is that as we go far enough back in time in the contracting universe we will reach a point, defined here as time t = $\rm{t}_0$, or just t = 0, (see figure \ref{fig:figure1}) where the degrees of freedom that our universe is made of may get replaced by other degrees of freedom \cite{Vafa&al}. Somewhat different kind of transition appears in the scenario of \cite{Rai_1, Rai_2}. At energy scale $\Lambda_{cr} \sim 10^{10} - 10^{16}$ GeV new topological degrees of freedom replace the standard model particles.  

We start with supersymmetric topological matter in the early phase I and to move towards phase II, where SUSY is a priori not guaranteed to exist. The fate of SUSY is determined at $t \sim 0$ when both time derivatives of $\rho^I_m$ and $\rho^{II}_m$ are non-zero, as in figure \ref{fig:figure1} green area. 

We assume no observables of the topological phase I will distinguish positions, so the metric should be homogeneous, i.e. a constant curvature metric. The time direction is picked out as an invariant concept in both phases. We would like to determine the consequences of this for the geometry in phase I as viewed from the frame II perspective. The most general metric with these symmetries is
\begin{equation}
	ds^2=-dt^2+a^2(t)\left[\frac{dr^2}{ (1-k r^2)} +  r^2 d\Omega^2\right]
	\label{FRWansatz}
\end{equation}
where $k=+1,0,-1$ for positive, flat or negative curvature spaces. As discussed in subsection \ref{brst}, the solutions to BRST \cite{BRS, Tyutin} invariant configurations in 4D topological gravity are conformally flat, self-dual geometries, which have zero Weyl tensor
\be
	W_{ABCD} = 0
\ee
This condition by itself allows all three possibilities for k. We will view time as a continuous element between phase I and phase II. Thus, a natural assumption is that the metric can be expressed as a flat metric up to a conformal factor that is only dependent on time, which is the only duality invariant coordinate. This is equivalent to having an FLRW metric \eqref{FRWansatz} with $k=0$
\begin{align}
	ds^2 = a^2(\eta)(-d\eta^2+ dx^i dx^i)
\end{align}

Physics in phase II after reheating is well described by a thermal distribution of SM matter (and the dark components). The notion of time is common to both phases of the universe. This leads to energy being common to both phases. In addition there are weak long range correlations that originate from phase I modes that are non-local in phase II. 

\section{Topological models in phase I}
\label{topolgrav}

\subsection{General properties of topological models}
\label{brst}

In topological models, the horizon problem is solved simply because the locality, relevant in our universe in phase II, is not natural in phase I \cite{Vafa&al}. The light modes of phase I are non-local as viewed from phase II. A known example are the winding modes of the string gas cosmology \cite{Brandenberger:2004}. Fluctuations visible in phase II are not part of the degrees of freedom of phase I.

\begin{figure}[H]
	\centering
	\captionsetup{width=.8\linewidth}
	\includegraphics[width=12cm]{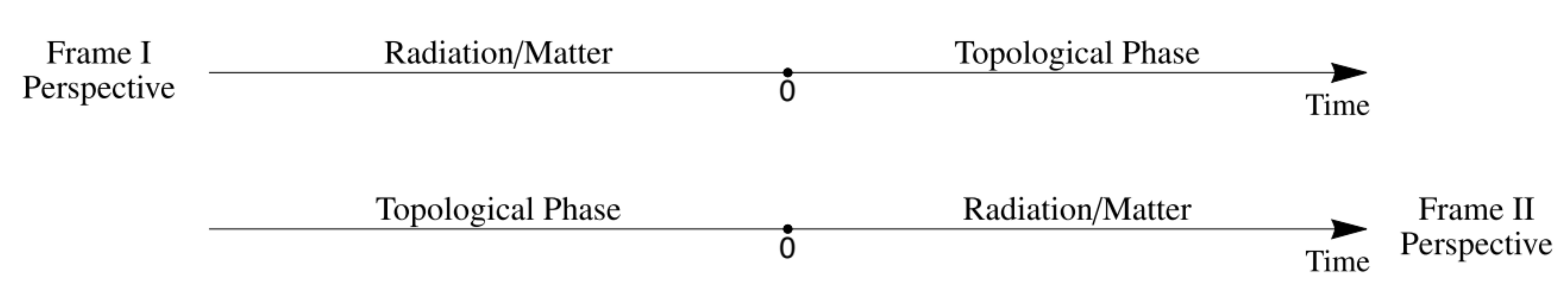}
	\caption{\small The degrees of freedom making up phase I are absent in a low energy description of phase II. Therefore the former appears topological from the point of view of the latter. This relation is also true with the roles of phase I and II interchanged.}
	\label{fig:figure2}
\end{figure}

How does phase I look from the perspective of phase II \cite{Vafa&al}? In phase I there should not be any position dependent observables. Let us assume the state in phase I is given by $| I\rangle$. We would expect $n$-point correlations of physical observables in this state
\be
\langle I | {\cal O}^{i_1}(x_1)\ldots {\cal O}^{i_n}(x_n)|I \rangle =A^{i_1,\ldots,i_n}
\label{correlations}
\ee
to be position independent when all $\partial_j A^{i_1,\ldots,i_n} = 0$. This is a key feature of a topological quantum field theory. While we view phase I as a topological phase from the perspective of frame II it is curious that the reverse is also true: phase II can be viewed from the perspective of frame I as a topological theory \cite{Vafa&al}. This is illustrated in figure \ref{fig:figure2}.

In topological field theories observables must be a measure of global features. Consequently, there are no propagating signals. This property is achieved in the Becchi-Rouet-Stora-Tyutin (BRST) \cite{BRS, Tyutin} formalism by the presence of a Grassmann odd charge operator $Q$.

This operator $Q$ is nilpotent, hermitian, and it commutes with the Hamiltonian, $[H,Q] = 0$. The action of the charge operator on fields $\Phi$ is given by 
\be 
\delta\Phi = i \epsilon [Q,\Phi] 
\label{deltaphi}
\ee
where $\epsilon$ is a Grassmann parameter, a supernumber that anticommutes with all other Grassmann variables. $Q$ is also the Noether charge for the BRST symmetry. The action combines together bosonic and fermionic fields in a way similar to the pairing in supersymmetric theories. Physical states in the Hilbert space are $Q$-cohomology classes: these states are $Q$-closed (i.e. $|\psi\rangle$ satisfying $Q|\psi\rangle=0$) modulo $Q$-exact (i.e. $|\psi\rangle$ such that $|\psi\rangle = Q|\chi\rangle$ for some $|\chi\rangle$). This latter requirement implies that the fermionic partners of bosonic fields are in fact ghosts so that all degrees of freedom cancel in the BRST sense.

If we assume that the vacuum is $Q$-invariant, then $Q$-exact operators have a vanishing expectation value $\langle [ Q,\mathcal{O} ]\rangle = 0$. In topological field theories, the energy-momentum tensor (given by the variation of the action with respect to the metric) is $Q$-exact, i.e. $T_{\alpha\beta} = \{Q, \lambda_{\alpha\beta}\}$ for some $\lambda$. This implies that the partition function is invariant under metric variations
\begin{align*}
	\delta Z &= \int \mathcal{D}\Phi e^{-S}\left(-\delta S\right)
	= - \int \mathcal{D}\Phi e^{-S}\{Q,\int \sqrt{g}\delta g^{\alpha\beta} \lambda_{\alpha\beta}\}\\
	&= -\langle \{Q,\int \sqrt{g}\delta g^{\alpha\beta} \lambda_{\alpha\beta}\} \rangle = 0
\end{align*}
provided the integration measure is BRST invariant.

\begin{figure}
	\centering
	\captionsetup{width=.8\linewidth}
	\includegraphics[width=3cm]{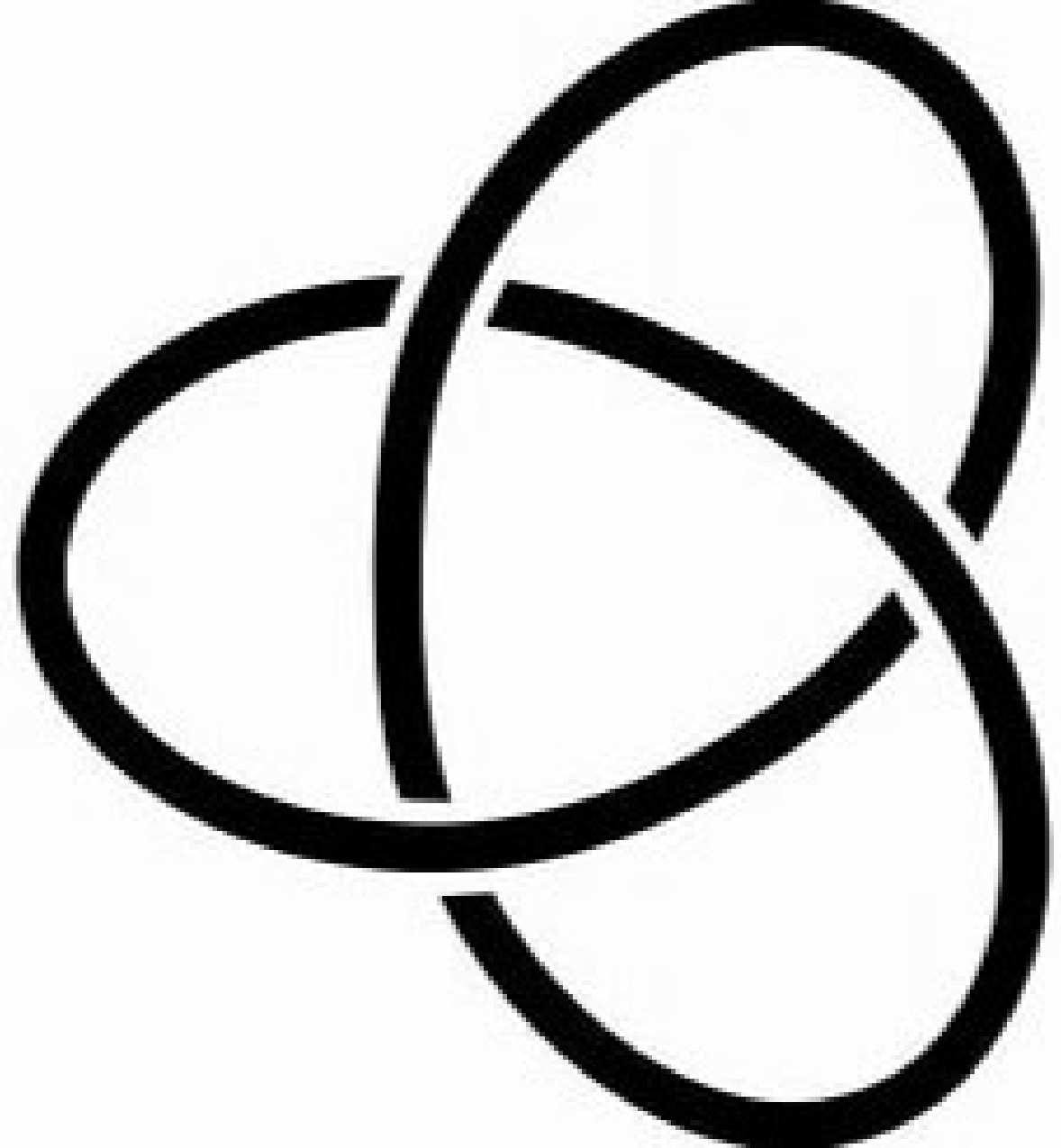}
	\caption{\small A trefoil knot in 3D space. The curve has orientation, clockwise or anticlockwise.}
	\label{fig:figure4}
\end{figure}

Another way to illuminate background independence in a topological theory in general is based on calculating Wilson loops in 3D Chern-Simons (CS) theory \cite{Witten_0}.\footnote{~CS theory is discussd later in section \ref{chernon}} Wilson loops give a natural class of gauge invariant observables that do not require a choice of metric. Let C be an oriented closed curve in M. Intrinsically C is simply a circle, but the topological classification of embeddings of a circle in M may be complicated, as we can imagine in figure \ref{fig:figure4}. Let R be an irreducible representation of G. One then defines the Wilson loop $W_R(C)$ to be the following functional of the connection $A_i$. One computes the holonomy of $A_i$ around C, getting an element of G that is well-defined up to conjugacy, and then one takes the trace of this element in the representation R. Thus, the definition is 
\be
W_R(C) = \rm{Tr}_R P \exp \int_C A_i dx^i
\ee
The crucial property of this definition is that there is no need to introduce a metric, so general covariance is maintained. 
Consider the partition function $Z$, defined as 
\be
Z = \int \mathcal{D}\mathcal{A}\exp(i\mathcal{L})\prod_i {W_R}_i (C_i)
\label{partitionfcn}
\ee
where $\mathcal{D}\mathcal{A}$ represents Feynman integral over all gauge orbits, the $C_i$ are non-intersecting knots and $R_i$ representation assigned to $C_i$. The partition function Z is thus automatically independent of any background metric. However, there is still a question of whether the theory contains local excitations.

\subsection{Fang and Gu's topological gravity}
\label{guth}

We consider the topological theory by Fang and Gu  \cite{Gu:2017, Gu:2021, Fang-Gu:2023}. The topological quantum field theory (TQFT) approach can not be easily generalized into 3+1D because consistency with Einstein's gravity in 3+1D contains propagating a mode, the graviton. Therefore it is obviously not a case for TQFT in the usual sense. Secondly, there is no Chern-Simons like action in 3+1D. Fang and Gu have shown that Einstein gravity might emerge by adding a topological mass term of the 2-form gauge field. Physically, such a phenomenological theory might describe a loop condensing phase, i.e. flux lines in the context of gauge theory.

Due to the recent developments in the classification of topological phases of quantum matter in higher dimensions \cite{Chenlong, Wenscience, cobordism, cobordisma, Wencoho}, new types of TQFT have been discovered in 3+1D to describe the  three-loop-braiding statistics. It is argued that such types of TQFT are closely related to Einstein gravity and that gravitational field will disappear at extremely high energy scale. 3+1D quantum gravity would be controlled by a TQFT renormalization group fixed point. At intermediate energy scales, Einstein gravity and classical spacetime would emerge via loop (flux lines) condensation of the underlying TQFT. The uncondensed loop-like excitation are a natural candidate of dark matter. Such kind of dark matter will not contribute scalar curvature but  will be a direct source of torsion. Normal matter, like Dirac fermions, will not contribute to torsion.

Let us begin with the topological gravity theory in 3+1D \cite{Topgravity}. Consider the following topological invariant action
\begin{eqnarray}
	S_{top}&=&\frac{k_{1}}{4\pi }\int \varepsilon
	_{abcd} R^{ab}\wedge e^{c}\wedge e^{d} +\frac{k_{2}}{2\pi }\int B_{ab} \wedge  R^{ab} \nonumber\\ &&+\frac{k_{3}}{2\pi }%
	\int \widetilde{B}_{a} \wedge  T^{a} 
\label{action}
\end{eqnarray}
where $e$ is the tetrad field, $R$ is the curvature tensor, $T$ is the torsion tensor and $B,\widetilde{B}$ are 2-form gauge fields. Like in the CS theory, the values of $k_i$ are quantized. Without loss of generality, the following values can be chosen $k_1=k_2=2$ and $k_3=1$ for convenience. The above action is invariant under the following (twisted) 1-form and 2-form gauge transformations, respectively:
\begin{eqnarray}
	e^{a} &\rightarrow &e^{a}+Df^{a} \nonumber\\ 
	B_{ab} &\rightarrow &B_{ab}-\frac{k_{3}}{2k_{2}}\left( \widetilde{B}%
	_{a}f_{b}-\widetilde{B}_{b}f_{a}\right)  \nonumber \\
	\widetilde{B}_{a} &\rightarrow &\widetilde{B}_{a}-\frac{k_{1}}{k_{3}}
	\varepsilon _{abcd}f^{b}R^{cd}, \label{1form}
\end{eqnarray}
and
\begin{eqnarray}
	B_{ab} & \rightarrow & B_{ab}+D\xi _{ab}, \label{2forma}\\
	\widetilde{B}_{a} &\rightarrow &\widetilde{B}_{a}+D \tilde{\xi} _{a} 
	\nonumber\\
	B_{ab} &\rightarrow &B_{ab}-\frac{k_{3}}{2k_{2}}\left( \tilde{\xi}  _{a}\wedge
	e_{b}-\tilde{\xi}  _{b}\wedge e_{a}\right).  \label{2formb}
\end{eqnarray}
Such an action can be regarded as the non-Abelian generalization of $AAdA+BF$ type TQFT \cite{aada1, aada2, aada3} of the Poincare gauge group. Physically, it has been shown that such kind of TQFT describes the three-loop-braiding statistics \cite{loop1, loop2}. As a TQFT, the action Eq. (\ref{action}) is a super-renormalizable theory. The coefficient quantization and canonical quantization of such a theory are discussed in \cite{Gu:2021}. 

SUSY generalization of 3 + 1D topological gravity is discussed in \cite{Gu:2017}. One needs to introduce the gauge connection of super Poincare group and write the action as $\int sTr[ A\wedge A\wedge (dA+A\wedge A)]+\int sTr(B \wedge F)$. For the $N=1$ case, one can express $A$, $B$ and $F$ as follows
\begin{eqnarray}
	&&A_{\mu}\equiv \frac{1}{2}\omega_{\mu}^{ab} M_{ab}+e_\mu^a P_a+\bar{\psi}_{\mu\alpha} Q^\alpha \nonumber\\
	&&B_{\mu\nu}\equiv \frac{1}{2}{B_{\mu\nu}}^{ab} M_{ab}+{\tilde{B}_{\mu\nu}}^a P_a+\mathfrak{B}_{\mu\nu\alpha} Q^\alpha \nonumber\\
	&& F_{\mu\nu}\equiv
	\frac{1}{2}{R_{\mu\nu}}^{ab} M_{ab}+T_{\mu\nu}^a P_a+\bar{R}_{\mu\nu\alpha} Q^\alpha
\end{eqnarray}
Here $\bar{R}_{\mu\nu\alpha}$ is the super curvature tensor defined as $\bar{R}_{\mu\nu\alpha}=D_{\mu} \bar{\psi}_{\nu\alpha}-D_{\nu} \bar{\psi}_{\mu\alpha}$ where $D_{\mu}$ is the covariant derivative for spinor fields. Fermionic loops (flux lines) cannot be condensed. Therefore supersymmetry breaking happens at very high energy scale when bosonic loops condense and classical space-time emerges. More details are presented in \cite{Fang-Gu:2023}.

Although the total action S is super-renormalizable, it does not imply UV-complete quantum gravity theory due to explicit breaking of 2-form gauge symmetries by the $S_\theta = -\frac{\theta}{2\pi} \int B_{ab}\wedge B^{ab}$ term. The algebraic tensor 2-category theory \cite{TCAT1,TCAT2} may provide an equivalent UV-complete description for a topological quantum gravity theory in 3+1D.

In \cite{Fang-Gu:2023} the authors give a more profound treatment. It includes a deformation parameter $\lambda$ which represents the bare cosmological constant term. It plays a crucial role in this scenario. $\lambda = 0$ corresponds to a trivial universe with vanishing Riemann curvature, while $\lambda \neq 0$ corresponds to a non-trivial universe where Einstein gravity arises at low energy. In this scenario SUSY does not survive at energy scale below Planck energy.

\subsection{Chern-Simons model in phase O}
\label{chernon}

We disclose arguments for preons. The distinctive feature of our preons (called here chernons) is the treatment of SUSY as unbroken global symmetry with the particles in supermultiplets. The chiral and vector supermultiplets for three colors are given in table \ref{tab:table1} \cite{Rai_1}.
\begin{table}[H]
	\begin{center}
		\captionsetup{width=.8\linewidth}
		\begin{tabular}{|l|l|} 
			\hline
			Multiplet & Particle, Sparticle \\ 
			\hline   
			chiral multiplets spins 0, 1/2 & 
			$s^-$, $m^-$;~a, n \\
			vector multiplets spins 1/2, 1 & $m^0$, $\gamma$;~$m_C, g_C$ \\  
			\hline
		\end{tabular}
		\caption{\small The particle $s^-$ is a neutral scalar particle. The particles $m^-, m^0$ are charged and neutral, respectively, Weyl spinors. The a is axion and n axino. $m^0$ is color singlet particle and $\gamma$ is the photon. $m_C$ and $g_C$ (C=R,G,B) are zero charge color triplet fermion and boson, respectively.}
	\label{tab:table1}
\end{center}
\end{table}  

Chernon interactions are 2+1 dimensional inside a 3+1D world. Chern-Simons-Maxwell (CSM) interaction models have been studied in condensed matter physics, e.g. \cite{Deser_J_T, Giro_G_d_M_N, Beli_D_F_H}. In this note we extrapolate the CSM model a long way to particle physics phenomenology at high energy in the early universe. 

We construct the visible matter of two fermionic chernons: one charged $m^-$, one neutral $m^0$, and the photon. The Wess-Zumino \cite{WZ} type  action \cite{Rai_1} is supersymmetric as well as C symmetric. The chernons have zero (or very small) mass. 
The chernon baryon (B) and lepton (L) numbers are zero. Given these quantum numbers, quarks consist of three chernons, as indicated in table \ref{tab:table4}.\footnote{There are more combinations of states like those containing an $m^+m^-$ pair. This state annihilates immediately into other chernons, which form later leptons and quarks.}

In \cite{Beli_D_F_H} a 2+1 dimensional Chern-Simons (CS) action \cite{cs, Witten_0} was used to derive chernon-chernon interaction, which turns out to trigger the fist phase transition between O and II. In 2+1 dimensions, a fermionic field has its spin polarization fixed up by the sign of mass \cite{Binegar, Deser-J, Jackiw-N, Frohlic-M}. The model includes two positive-energy spinors (two spinor families) and a complex scalar $\varphi$. The fermions obey Dirac equation, each one with one polarization state according to the sign of the mass parameter. 

The chernon-chernon scattering amplitude in the non-relativistic approximation is obtained by calculating the t-channel exchange diagrams of the Higgs scalar and the massive gauge field. The propagators of the two exchanged particles and the vertex factors are calculated from the action \cite{Beli_D_F_H}.

The gauge invariant effective potential for the scattering considered is obtained in \cite{Kogan, Dobroliubov}
\begin{equation}
	V_{{\rm MCS}}(r)=\frac{e^{2}}{2\pi }\left[ 1-\frac{\theta }{m_{ch}}\right]
	K_{0}(\theta r)+\frac{1}{m_{ch}r^{2}}\left\{ l-\frac{e^{2}}{2\pi \theta }%
	[1-\theta rK_{1}(\theta r)]\right\} ^{2} 
	\label{Vmcs}
\end{equation}
where $K_{0}(x)$ and $K_{1}(x)$ are the modified Bessel functions and $l$ is the angular momentum ($l=0$ in this note). In (\ref{Vmcs}) the first term $[~]$ corresponds to the electromagnetic potential, the second one $\{~\}^2$ contains the centrifugal barrier $\left(l/mr^{2}\right)$, the Aharonov-Bohm term and the two photon exchange term.

One sees from (\ref{Vmcs}) the first term may be positive or negative while the second term is always positive. The function $K_{0}(x)$ diverges as $x \ra 0$ and approaches zero for $x \ra \infty$ and $K_{1}(x)$ has qualitatively similar behavior. For our scenario we need negative potential between equal charge chernons. Being embarrassed of having no data points for several parameters in (\ref{Vmcs}) we can give one relation between these parameter values for a binding potential. We must require the condition\footnote{For applications to condensed matter physics, one must require $\theta \ll m_{e}$, and the scattering potential given by (\ref{Vmcs}) then comes out positive \cite{Beli_D_F_H}.}
\be
\theta \gg m_{ch}
\label{condition}
\ee
The potential (\ref{Vmcs}) also depends on $v^{2}$, the vacuum expectation value, and on $y$, the parameter that measures the coupling between fermions and Higgs scalar. Being a free parameter, $v^{2}$ indicates the energy scale of the spontaneous breakdown of the $U(1)$ local symmetry. 

A summary of the three phases and their properties is given in table \ref{tab:table2}.

\begin{table}[H]
	\begin{center}
		\captionsetup{width=.8\linewidth}
		\begin{tabular}{|l|l|l|l|l|} 
			\hline
			Ph. & HE particles & HE symm. & Low energy symm. \\ 
			\hline 
			I & F\&G theory  & SUSY & \cancel{SUSY} \\				 
			O & chernons & SUSY & SM; \cancel{SUSY} \\  		
			II & SM particles & SUSY GUT & SM; \cancel{SUSY}? \\
			\hline
		\end{tabular}
		\caption{\small Development of the universe from phase I to phase O and finally to phase II. The phase O's role is to hide supersymmetry, create SM matter, spacetime metric and baryon asymmetry in the universe. In the rightmost column SM stands for $SU(3) [\times SU(2)]\times U(1)$. The term $[\times SU(2)]$ indicates appearance of weak interaction "automatically" between u- and d-quarks as well as between e and $\nu$.} 
		\label{tab:table2}
	\end{center}
\end{table}

\section{Topological early phases versus inflation}
\label{compar}

\begin{figure}
	\centering
	\captionsetup{width=.8\linewidth}
	\includegraphics[width=14cm]{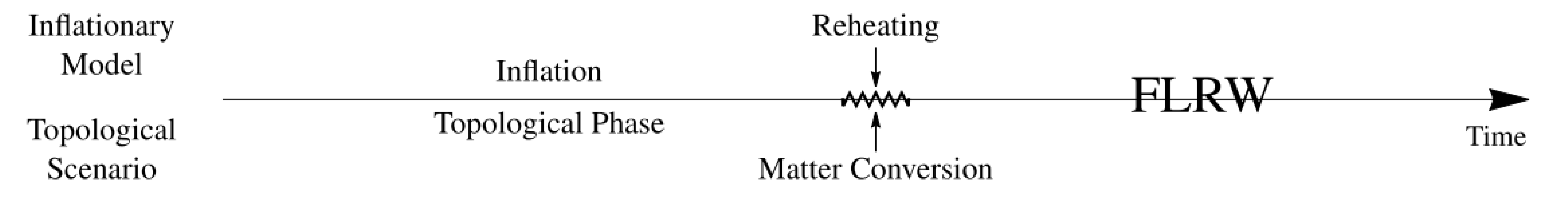}
	\caption{\small Comparison between the inflationary and topological paradigms for the early universe. The topological scenarios replaces the period of accelerated expansion by a topological phase to explain homogeneity, isotropy, flatness and near scale invariance. In both paradigms, the universe for $t > 0$ is well described by the standard Big Bang cosmology.}
	\label{fig:figure3}
\end{figure}

In this section we compare and contrast the topological scenarios with the inflationary scenario. There are a number of common features in the two approaches  as can be seen in Fig.~\ref{fig:figure3}.

The end result for both is the FLRW scenario.  Both of them involve a kind of phase transition. In the case of inflation the transition is marked by the end of the early expansion and beginning of reheating as the inflaton settles to the minimum of the potential. In the case of the topological scenario the phase transition takes place by a topology and symmetry change process \cite{Horowitz, Tanaka_N}. 
In both scenarios we have a nearly homogeneous thermal initial condition for FLRW in phase II. In both scenarios the homogeneity of space is described by a novel phenomenon: in the inflationary scenario by the exponential expansion of the space and in the topological phase by the fact that gravity is described by a topological theory. In the inflationary scenario the fluctuations of the inflaton field leads to scalar fluctuation, whereas in the topological phase which involves only global/zero modes and only through scale anomalies do we get fluctuations in the otherwise thermal background. Detailed properties and predictions of the topological inflation are presented in  \cite{Vafa&al}. Briefly said, processes take place as well as in other successful models. After reheating everything goes as in the standard model of cosmology.

\section{Conclusions}
\label{conclusions}

There are three possibilities for the fate of low energy supersymmetry: no SUSY at all, highly broken SM SUSY, and hidden SUSY (in chernons or in some other way). We consider the first case unlikely. The second case has been studied thoroughly with certain success but the SM superpartners are still missing. The third case, described above, agrees with the standard model particle spectrum and provides an answer to matter-antimatter asymmetry by the mechanism presented in \cite{Rai_2}. 

We conclude it is premature to consider supersymmetry a dream. Instead, a rich spectrum of light, laboratory observable bosonic and fermionic states are predicted by the supermultiplet table \ref{tab:table1} as colored constituents making singlet composites.

\newpage
\appendix
\section{Chernon-particle correspondence}
\label{appdx}

The matter-chernon correspondence for the two first flavors is indicated in table \ref{tab:table4}. 

\begin{table}[H]
	\begin{center}
		\captionsetup{width=.8\linewidth}
		\begin{tabular}{|l|l|} 
			\hline
			SM Matter 1st gen. & Chernon state \\ 
			\hline                                       
			$\nu_e$ & $m^0_R m^0_G m^0_B$  \\     
			$u_R$ & $m^+ m^+ m^0_R$  \\			  
			$u_G$ & $m^+ m^+ m^0_G$  \\  			
			$u_B$ & $m^+ m^+ m^0_B$  \\
			
			$e^-$ & $m^- m^-m^-$  \\		         
			$d_R$ & $m^- m^0_G m^0_B$ \\		 
			$d_G$ & $m^- m^0_B m^0_R$ \\			
			$d_B$ & $m^- m^0_R m^0_G$ \\
			\hline
			W-Z Dark Matter & Particle \\
			\hline   
			boson (or BC) & $s^0_r$, axion(s) \\
			$e'$ & axino $n$ \\
			meson, baryon $o$ & $n\bar{n}, 3n$ \\
			nuclei (atoms with $\gamma ')$ & multi $n$ \\
			celestial bodies & any dark stuff \\	 
			black holes & anything (neutral) \\
			\hline
		\end{tabular}
		\caption{\small Visible and Dark Matter with corresponding particles and chernon composites. 
		$e'$ and $\gamma'$ refer to dark electron and dark photon, respectively. BC stands for Bose condensate. Chernons obey anyon statistics. The binding of chernon composites is described in section \ref{chernon}.}
		\label{tab:table4}
	\end{center}
\end{table}


\end{document}